\documentclass{elsart}
\input{epsf.tex}

\usepackage{amssymb}

\begin{document}
\newcommand{\be}{\begin{equation}}
\newcommand{\nn}{\nonumber}
\newcommand{\ee}{\end{equation}}
\newcommand{\bea}{\begin{eqnarray}}
\newcommand{\eea}{\end{eqnarray}}
\newcommand{\wee}[2]{\mbox{$\frac{#1}{#2}$}}   
\newcommand{\unit}[1]{\,\mbox{#1}}
\newcommand{\degree}{\mbox{$^{\circ}$}}
\newcommand{\ltish}{\raisebox{-0.4ex}{$\,\stackrel{<}{\scriptstyle\sim}$}}
\newcommand{\vs}{{\em vs\/}}
\newcommand{\bin}[2]{\left(\begin{array}{c} #1 \\
#2\end{array}\right)}
\newcommand{\p}{_{\mbox{\small{p}}}}
\newcommand{\m}{_{\mbox{\small{m}}}}
\newcommand{\tra}{\mbox{Tr}}
\newcommand{\rs}[1]{_{\mbox{\tiny{#1}}}}        
\newcommand{\ru}[1]{^{\mbox{\small{#1}}}}

\begin{frontmatter}



\title{Strong causality from weak via continuous monitoring} 
\author[label1]{John Jeffers}
\address[label1]{SUPA, Department of Physics, University of Strathclyde, John Anderson Building, 107 Rottenrow, Glasgow G4 0NG, UK}
\ead{john@phys.strath.ac.uk}

\begin{abstract}
Repeated unbiased measurements cause a continual application of the weak causality principle, leading to an apparent arrow of time for continuously-monitored quantum systems.
\end{abstract}

\end{frontmatter}

\section{Introduction}

There seems to be a natural temporal order of events - an order that has not been found in the laws of physics. This conflict between our everyday world, in which every effect has a prior cause, and the equations, which display temporal symmetry, has been the subject of much research \cite{halliwell96,zeh07}. The everyday asymmetry can sometimes be traced to different boundary conditions imposed on physical systems in the future and the past. A simple example would be the second law of thermodynamics, which is thought to originate from a very smooth early state of the universe. As has been pointed out \cite{pegg06}, quantum mechanics is so fundamental that there have been attempts to link a temporal arrow to the theory \cite{halliwell96,zeh07,schulman97,savitt97}. Evolution in quantum theory is unitary and time-symmetric, and so the linking has concentrated on the measurement process, which at least has the veneer of nonunitarity. 

Two types of causality have been identified by some authors \cite{cramer80}. A macroscopic cause is one which can be initiated by an observer, such as by preparing a system in a particular state. Similarly, a macrosopic effect is one which allows an observer to receive information. All other causes and effects are microscopic. We normally think of an such an observer as an agent which can put information into, or extract it from a quantum system. More generally, however, we can consider anything which prepares or measures the state of a quantum system as an observer. Typically the distinction between microscopic and macroscopic is not such a sharp one, but the criterion employed here allows a sharp divide based on information, and not directly on system size. The ensuing weak causality principle is based on this divide. This principle requires macroscopic causes to precede their macroscopic effects in any reference frame. Thus weak causality merely prevents information from being sent to an observer in the past. Such information cannot travel backwards in time, or outside the light cone. Strong causality, however, implies that cause always precedes effect, even microscopically. EPR-type quantum correlations appear to violate strong causality, but not weak causality, as they do not result in macroscopic information transfer between spacelike-separated systems.  

It has been shown \cite{pegg06} that quantum mechanics does not impose strong causality on a system. Furthermore the weak causality that the theory does satisfy does not follow from quantum theory. Rather it is imposed by the condition that the measurements should be unbiased - that every state in an orthogonal set spanning the system space be equally represented in the set of possible measurement results. The consequences of this are explored here for systems which are subjected to repeated rapid measurements. 

In this paper it is pointed out that open quantum systems can be viewed as being subject to a rapid sequence of such measurements. The standard unitary evolution is punctuated by a series of measurement events, each of which is assumed to occur on a fast (effectively instantaneous) timescale. These events each cause a slight change of the state via the projective Kraus effect operators - the modern description of the effect of Von Neumann's collapse. Any measurement process which occurs is not described by this formalism, but at each measurement weak causality is imposed by the measurement boundary conditions. These are the only properties of the measurement process that are explicitly used here. In the limit where the weak measurements occur in rapid succession an individual measurement makes almost no change to the state. However, their collective effect is that the system has causality imposed upon it quasi-continuously. 

The paper is organised as follows. In section \ref{causmeas} the implications for causality of the unbiased measurement condition inherent in conventional quantum mechanics are briefly described. This topic is explored in greater depth elsewhere \cite{pegg06,pegg02b,pegg05}. Section \ref{MME}, provides an overview of continuous monitoring and its description in terms of master equations and trajectory solutions. Section \ref{causMME} describes the application of the unbiased measurement condition to continuously-monitored systems, and in the final section conclusions are provided.

\section{Causality and measurement in quantum systems}
\label{causmeas}

\subsection{Quantum mechanical probability postulate}

Suppose that a quantum system is prepared in a state selected from a set of unit trace density operators $\{\hat{\rho}_i\}$ with prior probability $P_i$. The system is later measured. Measurement is conventionally described by a set of probability operators $\{\hat{\pi}_j\}$, one for each possible measurement outcome $j$ \cite{helstrom76}. When the measurement result $j$ is obtained the system is measured to be in the state $\hat{\pi}_j/ \tra \hat{\pi}_j$. The denominator $\tra \hat{\pi}_j$ is simply the probability that the measurement result $j$ would be obtained if the system were initially prepared in the maximally mixed state - one which is proportional to the unit operator of the system state space $\hat{1}$. The standard requirement that a measurement of the system always gives one of the results $j$ leads to the condition 
\bea
\label{unbcond}
\sum_j \hat{\pi}_j =\hat{1}, 
\eea
so that the trace of this sum is the dimension of the system state space. This condition has another consequence, however. It imposes a lack of bias on the measurement: all states from an orthogonal set spanning the system space are equally represented, and the measurement gives a true reading of the prepared state, within the limits imposed by quantum mechanics. Note that there is no such condition which applies to the possible prepared states, which are unresticted. The standard probability postulate of quantum mechanics is written in terms of the prepared state and the probability operators. The probability that measurement result $j$ is obtained given that the preparation outcome was $i$ is
\bea
\label{born}
P(j|i) = \tra \left( \hat{\rho}_i \hat{\pi}_j \right).
\eea
The different trace properties of preparation and measurement operators ensure that the conventional probability postulate is asymmetric in its treatment of preparation and measurement. Preparation is biased but measurement is not. 

Suppose instead that the measurement device is biased; it does not treat all states from an orthogonal set spanning the space equally (for example, by simply disregarding some of them with a certain probability, or by being less sensitive to some measurement outcomes than others). It is then normally necessary to use a symmetric form of quantum mechanics, which treats both preparation and measurement equally \cite{pegg02b,pegg02a}. However, it has been shown that a useful prepared density operator, which depends solely on preparation device properties, can only be suitably assigned to the system when the measurement device is unbiased. The measurement in this case can be described by probability operators which satisfy eq. (\ref{unbcond}). The symmetric formalism becomes equivalent to the standard asymmetric formalism and eq. (\ref{born}) applies.

The probability $P_i$ is calculated from system properties, and if it is to be a probability of preparing a state it should depend on preparation properties only. If the measurements are biased, however, it depends on both preparation and measurement device properties. The only way for the measurement properties to have no influence is for the measurement to be unbiased, so that no information from the measurement affects $P_i$. Given that measurements occur after preparation this means that unbiased measurement, via its mathematical expression in eq. (\ref{unbcond}), imposes weak causality on the system \cite{pegg05,pregnell}. Weak causality is ``inserted {\it into} quantum mechanics, rather than arising from it" \cite{pegg06}. The above argument is introduced in \cite{pregnell} and appears in summarised form in section 2 of \cite{pegg05} and more extensively in \cite{pegg06}. 

\subsection{Evolution}
\label{evolution}
If evolution between preparation and measurement is included it is typically assumed that the prepared state evolves forward in time to the measurement time, in accordance with the principle of strong causality. Quantum systems do not exist in isolation, but interact with their surroundings. This interaction is normally described using a master equation, and the resulting evolution shows decoherence, in which quantum correlations are damped out and become classical probabilities. Most open systems are studied using a master equation which has Lindblad form \cite{lindblad76} as this implies that the system is Markovian, and that the environment does not feed lost information back into the system. One general form of the master equation is
\bea
\label{lindme}
\dot{\hat{\rho}} = -\frac{i}{\hbar} \left[ \hat{H},\hat{\rho} \right]+ \gamma \left( \sum_k \hat{b}_k \hat{\rho} \hat{b}^\dagger_k - \frac{1}{2}\hat{b}^\dagger_k \hat{b}_k \hat{\rho} - \frac{1}{2} \hat{\rho} \hat{b}^\dagger_k \hat{b}_k \right),
\eea
where $\hat{b}_k$ is a system operator and $\gamma$ is a rate, typically a decay. Any ``collapse of the wavefunction" occurs at the measurement time, after all of the evolution has taken place. A master equation of this form can be used to describe both single system evolution, in line with the above, or an ensemble of such systems. 

An individual system which undergoes nonunitary evolution described by a master equation of the above form can also be modelled in a different way, using the quantum trajectory method \cite{carmichael83}. The system is thought of as being in a pure state, evolving continuously in Hilbert space, but undergoing a random sequence of quantum jumps. This forms what is known as a trajectory or realisation for the system. Because of the random stochastic nature of the jumps there are many possible realisations. When these realisations are suitably averaged, the evolution obtained is identical to that of the Lindblad master equation. The Lindblad master equation gives the average behaviour of the quantum trajectories. 

This picture is not the only possible one, however. The system can also equally-well be described by the measured state evolving (and decohering) backwards in time towards the preparation time according to the retrodictive master equation \cite{barnett01}, or more simply by the evolution equation for the probability operators 
\bea
\label{rme}
\dot{\hat{\pi}} = -\frac{i}{\hbar} \left[ \hat{H},\hat{\pi} \right]- \gamma \left( \sum_k \hat{b}^\dagger_k \hat{\pi} \hat{b}_k - \frac{1}{2}\hat{b}_k \hat{b}^\dagger_k \hat{\pi} - \frac{1}{2} \hat{\pi} \hat{b}_k \hat{b}^\dagger_k \right).
\eea
In this picture the collapse occurs at the preparation time. Alternatively the system can even be described by both the prepared and measured states evolving in their respective time-directions to some intermediate collapse time between the preparation and measurement time. In each case all probabilities calculated are identical \cite{pegg06,pegg02b,pegg02a,barnett01}, and weak causality is not violated. Note that evolution under the retrodictive master equation does not correspond to recoherence. Information about the measured state is lost as time decreases, until the moment of collapse at the preparation time. For dissipative systems, however, which lose energy in the forward time direction, the retrodictive master equation corresponds to energy gain, backwards in time \cite{barnett00,jedrkiewicz04}. Time evolution in either direction appears to be equally-valid, and the system does not exhibit strong causality. Only weak causality applies, and even that is externally imposed.

\section{Continuous monitoring}
\label{MME}
Decoherence due to the continuous interaction with an unmeasured environment is not the only possible open system evolution. Measurement itself provides another possible type of environmental interaction, one which can cause discontinuous jumps in the state of the system. If the measurement is performed repeatedly, and randomly in time, the system state will undergo an evolution akin to a quantum trajectory. A sequence of measurements performed on an ensemble of such systems allows a description of the average system behaviour in terms of a Lindblad master equation \cite{caves87,milburn96,cresser06,hornberger07}. For a single system and for measurement results which correspond to discrete system variables, after a measurement the state of the system changes to 
\bea
\label{effect}
\hat{\rho} \rightarrow \hat{\rho}_{kl} = \frac{\hat{A}_{kl} \hat{\rho} \hat{A}^\dagger_{kl}}{\tra \left( \hat{\rho} \hat{A}^\dagger_{kl}\hat{A}_{kl} \right) },
\eea
where $\hat{A}_{kl}$ is the effect operator corresponding to measurement result $k$, and $l$ is an index which allows for the fact that each measurement result can be associated with more than one effect \cite{kraus83}. Effect operators not only provide the measurement result probabilities, but also describe the outcome of the evolution which the measurement imposes on the system. For example, the effect operator $\hat{A}_{1}=|0\rangle \langle 1|$, where $|n\rangle$ is the $n$-photon number state for a single mode of the electromagnetic field, corresponds to the detection of a single photon by a perfect absorbing photodetector. If the field has one photon excited the density operator is $\hat{\rho}_1=|1\rangle \langle1|$. Then the effect operator changes the state to $\hat{A}_{1}\hat{\rho}_1\hat{A}_{1} = |0\rangle \langle0|$. The one-photon field state becomes the vacuum state, and no further photons can be detected. 

The utility of the effect operator formalism is obvious, especially in that it avoids describing the measurement process. Typically it is thought that the system of interest becomes entangled with a measurement device which then performs some kind of projection on the state. The formalism naturally includes projections made on one part of an entangled nonlocal system. It also dovetails with the probability operator description of measurements, as the probability operator corresponding to measurement result $k$ can be written in terms of the effect operators as 
\bea
\hat{\pi}_k= \sum_l \hat{A}^\dagger_{kl}\hat{A}_{kl},
\eea
so the denominator in eq.(\ref{effect}) is the probability that the measurement result is $k$. If the measurement result is not known the evolution will be a weighted sum of all of the possible measurement-induced evolutions, and using eq. (\ref{unbcond}) it is possible to derive a master equation with a simple Lindblad form \cite{lindblad76},
\bea
\label{discretevar}
\dot{\hat{\rho}} = -\frac{i}{\hbar} \left[ \hat{H},\hat{\rho} \right]+ R \left( \sum_{k,l} \hat{A}_{kl} \hat{\rho} \hat{A}^\dagger_{kl} -\hat{\rho} \right). 
\eea
Here $R$ is the average rate of measurements, each of which is assumed to occur instantaneously. The strength of the measurements - how much disturbance they make on the system - is governed by the exact form of the effect operators. The measurement strength and rate may be combined to form an overall decay parameter such as that in eq. (\ref{lindme}), depending on the details of the system. 

An analogous master equation appropriate for monitoring by measurements which correspond to continuous variables can be derived \cite{barnett05,barnett06},
\bea
\label{ctsvar}
\dot{\hat{\rho}} = -\frac{i}{\hbar} \left[ \hat{H},\hat{\rho} \right]+ R \left( \int d\xi \int dx \int dp \hat{A}_\xi(x,p) \hat{\rho} \hat{A}^\dagger_\xi (x,p) -\hat{\rho} \right). 
\eea
where $\xi$ plays the role of the parameter $l$ in the discrete equation, and $x$ and $p$ are, say, position and momentum, so the effect operators correspond to joint measurements of both the position and momentum of the system \footnote{In \cite{caves87} and \cite{milburn96} similar equations were provided but for measurements of position {\it or} momentum. Joint measurements are important if the system is to have a good classical limit. This problem was circumvented in \cite{caves87} by introducing an extra feedback to cancel the effect of jumps.}. This type of weak joint monitoring can be caused by scattering, but in quantum optics the measurements could equally-well correspond to monitoring the real and imaginary parts of a coherent amplitude. In the appropriate limit of weak rapidly-repeated measurements a Lindblad master equation is obtained which is similar to those describing decoherence \cite{barnett05,barnett06}. This monitoring approach to scattering has been used by others \cite{hornberger07}, and similarly, the relations between scattering, monitoring, records and decoherence have been explored, typically in the context of quantum Brownian motion or friction \cite{joos85,gallis90,hornberger03,diosi95,vacchini00,dodd03}. Similar equations to (\ref{discretevar}) have also been studied in relation to continuous quantum walks \cite{kendon07}

Quantum trajectory analysis \cite{carmichael83} of such equations provides individual possible realisations of sequences of measurement results and, via the effect operators, possible measurement-driven evolutions. These realisations, when averaged over, provide the same results as the ``average behaviour" master equations. Each trajectory itself typically consists of smooth evolution punctuated by discontinuous jumps caused by measurement. For continuous monitoring master equations trajectories represent sequences of events which could actually occur. The discontinuous jumps associated with measurement events are an integral part of the evolution under the measurement master equation. Smooth average solutions are not always so useful, as they do not necessarily describe an evolution which could occur in practice, as shown in \cite{cresser06}. This distinguishes the measurement master equation from those which describe decoherence and other smooth (although short timescale) effects, and for which the quantum trajectory approach is more of an interpretation. 

\section{Causality and continuous monitoring}
\label{causMME}

\subsection{Repeated measurements and evolution} 
\begin{figure}[!htb]
\centerline{ \epsfxsize=100mm \epsffile{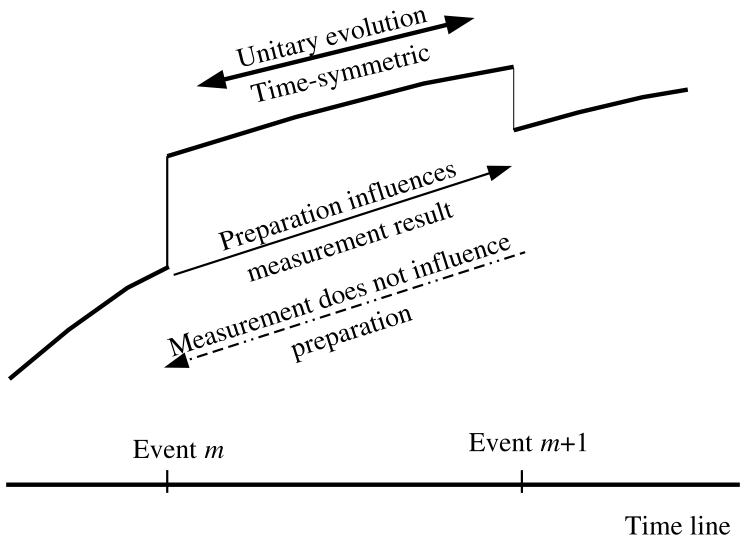}}
\caption{Schematic realisation of a trajectory in the neighbourhood of two successive measurement events. The measurement boundary condition stops measurement $m+1$ influencing the earlier preparation $m$, formed from the effect operators associated with the measurement $m$}
\label{fig1}
\end{figure}

Consider a single quantum trajectory corresponding to a possible (real) realisation of a repeatedly-measured system, such as in Fig. \ref{fig1}. 
When a measurement occurs, provided that no results are excluded, the unbiased measurement condition applies, and weak causality is imposed on the system. The unbiased measurement criterion is an integral component of eqs. (\ref{discretevar}) and (\ref{ctsvar}), so any system governed by them has weak causality repeatedly imposed upon it. Furthermore, through the effect operators, each measurement acts as a preparation device for the subsequent state. Each successive measurement guides the evolution of the subsequent state and later measurement results. It is unable to do this for prior measurement results because of the imposed causality condition. The evolution between two successive measurement events can be in either time direction, but any individual measurement event can only influence subsequent events. Thus measurement punctuates the `eventless' evolution by providing a sequence of events which can affect other such events in one time direction only. Backward influence is proscribed. 

\subsection{Retrodictive continuous monitoring?}

The argument presented above appears to leave little room for bi-directional temporal evolution of trajectories of continuously-monitored systems. It is easy, however, to derive a reverse-time evolution equation governing continuously-monitored systems by requiring that the probability $P(j|i)$ in eq. (\ref{born}) is independent of some hypothetical collapse time $t_c$ between the preparation and measurement times, $t_p$ and $t_m$ \cite{barnett01}. In standard quantum theory wavefuction collapse is assumed to occur at the measurement time, but in retrodictive quantum theory it occurs at the preparation time. For the purposes of this derivation it is assumed that the prepared state evolves forward in time and that the probability operator evolves backwards in time towards the collapse time. The invariance criterion is
\bea
\nonumber \frac{\partial }{\partial t_c} P(j|i) &=& 0\\
\nonumber \Rightarrow \tra \left( \frac{\partial \hat{\rho}_i}{\partial t_c} \hat{\pi}_j \right) &=& -\tra \left( \hat{\rho}_i \frac{\partial \hat{\pi}_j}{\partial t_c} \right)\\
\Rightarrow \tra \left( \frac{\partial \hat{\rho}_i}{\partial (t_c-t_p)} \hat{\pi}_j \right) &=& \tra \left( \hat{\rho}_i \frac{\partial \hat{\pi}_j}{\partial (t_c-t_m)} \right).
\eea
Insertion of eq. (\ref{discretevar}) and the cyclic property of the trace gives the backward time evolution equation of the probability operator
\bea
\label{rmme}
\frac{\partial \hat{\pi}_j} {\partial (-t)} = \frac{i}{\hbar} \left[ \hat{H},\hat{\pi} \right]+ R \left( \sum_{k,l} \hat{A}^\dagger_{kl} \hat{\pi}_j \hat{A}_{kl} - \hat{\pi}_j \right).
\eea
This equation is the continuous monitoring equivalent of eq. (\ref{rme}), and describes the evolution backwards in time of the probability operator corresponding to some final measurement on the system. Again, no calculated probability is affected by using this equation to describe the evolution of the system rather than eq. (\ref{discretevar}), so it appears that strong causality is not imposed. There is, however, a difference between this equation and its forward-time partner in that it is only useful for calculating probabilities or providing ensemble averages. In particular it does {\it not} describe measurements which occur backwards in time, as the effect operators appear in a different order. Measurements are not reversible if they provide any information about the state \cite{mensky96}. In terms of the effect operators the unbiased measurement criterion is $\sum_{kl} \hat{A}^\dagger_{kl} \hat{A}_{kl} = 1$, but for the equation above to describe time-reversed unbiased measurements the relation would have to hold for the operator order reversed. In general
\bea
\sum_{kl} \hat{A}_{kl} \hat{A}^\dagger_{kl} \neq 1
\eea
and the evolution which is described by the retrodictive equation corresponds to measurements made in the forward time direction. Only forward time trajectories correspond to possible measurement-driven evolutions. 

\subsection{Weak, rapid measurement limit}

As has been described, each measurement in a sequence has two purposes: it firstly imposes a condition on the system which in turn imposes weak causality, and secondly, it acts as a preparation device, creating the state which is subsequently measured. However, between each preparation and subsequent measurement only weak causality applies, and time reversal symmetry is still present. 

\begin{figure}[!htb]
\centerline{ \epsfxsize=100mm \epsffile{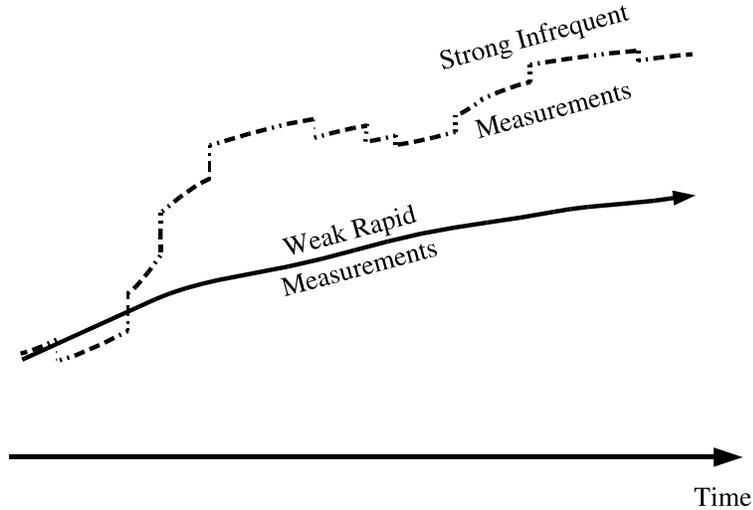}}
\caption{In the weak rapid measurement limit each measurement effects a miniscule collapse on the system, which can drive the evolution, and which repeatedly imposes causality, providing an apparent direction for time evolution.}
\label{fig2}
\end{figure}

If we make the measurements sufficiently rapidly, however, this effect will not be readily apparent in trajectory solutions corresponding to possible evolutions. This limit needs to be balanced by a weakening of the strength of the measurements, expressed mathematically for each effect operator as
\bea
\hat{A} \rightarrow \hat{1}
\eea
If the measurement strength is weak enough the evolution is neither solely driven by the measurements, nor suppressed by them so that Zeno-type effects occur \cite{streed}. Furthermore, in this limit trajectory solutions do not show discontinuous jumps, but correspond to smooth evolution of the state, as in Fig. \ref{fig2}. Whatever the measurement result the effect operator does not cause a large change in the state. This can be guaranteed by ensuring that the overall decay constant such as that in eq. (\ref{lindme}), which depends on both the strength and the rate, remains the same (see for example \cite{cresser06,barnett05,barnett06}). 
The details of the limit are system-dependent, but it should be noted that the measurements never become sufficiently-weak that they do not remove {\it any} information from the system, so the limit is never reached. However, in systems where the measurements are of discrete variables, such as monitoring the state of a two-level atom, this rapid weak measurement limit will lead to a simple Hamiltonian evolution of the system \cite{cresser06}. Conversely, in continuous-variable systems the weak measurement limit corresponds to states whose energy decays in time, as in systems governed by friction \cite{barnett05,barnett06,joos85,gallis90,hornberger03,diosi95,vacchini00,dodd03}. This familiar classical limit emerges, both in the expectation values, and in the rapid localisation of the state \cite{bellomo07}. The rapidity of the weak measurements imposes a quasi-continuous weak causality on the now continuously-evolving trajectory solution, from which the strongly-causal evolution emerges (Fig. \ref{fig2}). In this continuous limit the unbiased measurement criterion acts like a continuously-imposed boundary condition which guides the evolution forwards. 

\section{Conclusions}

The known laws of physics are time-symmetric, but events appear to occur in a definite order, and reconciling these facts is sometimes difficult. This paper uses the notion that measurement imposes weak causality on a quantum system to show that being monitored by its environment imposes causality upon the system quasi-continuously. 

The principle of weak causality, in this case expressed as the fact that a measurement cannot affect the prepared state, is equivalent to the familiar condition that a measurement should always have a result. If the condition is repeatedly applied, it provides a definite order for temporally-separated events (measurement outcomes). 

Any quantum system, e.g. an atom or molecule (or a small group of such), within a larger system repeatedly undergoes weak joint measurements of a pair of its conjugate variables, via collisions with the external system. Typically collisions will make a joint measurement of position and momentum \cite{barnett05,barnett06}. Scattering light off the system has the same effect. When we see an object we make repeated weak measurements of many of the subsystems which make up the object. In both of these cases a continuous monitoring master equation such as eq. (\ref{ctsvar}) will describe the system well. The rapidity with which such measurements are made on an open quantum system imposes a quasi-continuous weak causality on the system, leading to the appearance of strong causality in which all effects have definite causes that precede them. Thus if the classical world is regarded as a continuously monitored quantum system, then an important consequence of this paper is that there is no need to impose the strong causality principle on quantum mechanics.  Only weak causality is required to regain the strong causality of the classical world.  This allows more freedom to explore retrocausal interpretations of quantum mechanics, those that violate strong but not weak causality.

Note that the viewpoint expressed here does not depend on any particular measurement model, nor on any particular solution to the measurement problem. It does not require the process of measurement to be intrinsically time-asymmetric. All that is required is the measurement condition, which ensures that a result is obtained. Any model of measurement should satisfy this, unless some results are deliberately and artificially discarded. Of course, in most laboratory quantum physics experiments results are discarded, e.g. in quantum optical experiments results are often conditioned on the fact that a photodetector fires. This type of conditioning does not occur for environmental monitoring. 

Finally, it is customary to link arrows of time, hopefully showing that they point in the same direction. The unbiased measurement condition ensures that no probabilistic information from the measuring device, other than the states that can be measured and their relative weights, can enter the system. In other words these measurement quantities should be the same for all possible prepared states. For the type of monitoring proposed here the measuring device is the surroundings of the system. The continuously-applied ``boundary condition" implies that information does not flow from the surroundings into the system, and so is in accordance with the second law of thermodynamics. Conversely, it could be argued that for open systems the second law causes the unbiased measurement condition, and so this arrow of time would be simply a manifestation of the second law as it applies to measurement.

\section*{Acknowledgments}
I would like to thank David Pegg for useful discussions, and suggestions for improvements to the manuscript.

\end{document}